\documentclass[11pt]{article}

\usepackage[margin=1in]{geometry}
\usepackage{amsmath,amssymb}
\usepackage{graphicx}
\usepackage{booktabs}
\usepackage[authoryear,round]{natbib}
\usepackage{setspace}
\usepackage{caption}
\usepackage[hidelinks]{hyperref}
\usepackage{enumitem}

\setlist{itemsep=2pt,topsep=4pt}

\newcommand{\nQual}{256}
\newcommand{\minAB}{300}
\newcommand{\identityErr}{1e-16}
\newcommand{\nZeroHR}{3}
\newcommand{\powerName}{Jonathan Aranda}
\newcommand{\babipName}{Miguel Andujar}

\newcommand{\judgeC}{0.704}
\newcommand{\judgeBA}{0.331}
\newcommand{\judgeH}{0.139}
\newcommand{\judgeB}{0.384}

\newcommand{\rch}{-0.53}
\newcommand{\rchLo}{-0.61}
\newcommand{\rchHi}{-0.43}
\newcommand{\rcb}{-0.12}

\newcommand{\rhb}{-0.09}

\newcommand{\rcf}{-0.40}
\newcommand{\vc}{0.0072}
\newcommand{\vf}{0.0095}
\newcommand{\twocov}{-0.0066}
\newcommand{\vba}{0.0102}
\newcommand{\aHR}{0.0029}
\newcommand{\aB}{0.0080}
\newcommand{\aCov}{-0.0008}

\newcommand{\aGap}{5.2\%}
\newcommand{\aHRshare}{29\%}
\newcommand{\aBshare}{80\%}
\newcommand{\aCovshare}{-8\%}
\newcommand{\taHRshare}{55\%}
\newcommand{\taBshare}{65\%}
\newcommand{\bNoiseCur}{66\%}
\newcommand{\hNoiseCur}{21\%}
\newcommand{\nPairs}{19}
\newcommand{\medPairN}{183}
\newcommand{\medC}{0.86}
\newcommand{\minC}{0.81}
\newcommand{\maxC}{0.88}
\newcommand{\medH}{0.74}
\newcommand{\minH}{0.66}
\newcommand{\maxH}{0.80}
\newcommand{\medB}{0.41}
\newcommand{\minB}{0.24}
\newcommand{\maxB}{0.52}
\newcommand{\medBA}{0.44}
\newcommand{\minBA}{0.30}
\newcommand{\maxBA}{0.53}
\newcommand{\nBoth}{176}
\newcommand{\rCtwo}{0.82}
\newcommand{\rHtwo}{0.73}
\newcommand{\rBtwo}{0.28}
\newcommand{\muC}{1.211}
\newcommand{\muClo}{1.164}
\newcommand{\muChi}{1.258}
\newcommand{\pmuC}{0.771}
\newcommand{\sigC}{0.369}
\newcommand{\sigClo}{0.335}
\newcommand{\sigChi}{0.407}
\newcommand{\muH}{-3.059}
\newcommand{\muHlo}{-3.128}
\newcommand{\muHhi}{-2.989}
\newcommand{\pmuH}{0.045}
\newcommand{\sigH}{0.494}
\newcommand{\sigHlo}{0.439}
\newcommand{\sigHhi}{0.556}
\newcommand{\muB}{-0.853}
\newcommand{\muBlo}{-0.870}
\newcommand{\muBhi}{-0.835}
\newcommand{\pmuB}{0.299}
\newcommand{\sigB}{0.078}
\newcommand{\sigBlo}{0.053}
\newcommand{\sigBhi}{0.100}
\newcommand{\rhoCH}{-0.67}
\newcommand{\rhoCHlo}{-0.75}
\newcommand{\rhoCHhi}{-0.57}
\newcommand{\rhoCB}{-0.22}
\newcommand{\rhoCBlo}{-0.45}
\newcommand{\rhoCBhi}{0.00}
\newcommand{\rhoHB}{-0.23}
\newcommand{\rhoHBlo}{-0.48}
\newcommand{\rhoHBhi}{0.02}
\newcommand{\relC}{0.91}
\newcommand{\tauC}{0.114}
\newcommand{\relH}{0.76}
\newcommand{\tauH}{0.285}
\newcommand{\relB}{0.30}
\newcommand{\tauB}{0.121}
\newcommand{\mnAbsDC}{0.005}
\newcommand{\mnAbsDH}{0.005}
\newcommand{\mnAbsDB}{0.018}

\newcommand{\judgeBs}{0.308}
\newcommand{\babipShr}{0.326}
\newcommand{\mnAbsBorrow}{0.004}
\newcommand{\judgeBuni}{0.325}
\newcommand{\judgeBmulti}{0.309}
\newcommand{\nHist}{20,357}

\newcommand{\nFourHundred}{13}
\newcommand{\lastFourHundred}{1941}
\newcommand{\sdBAnineteen}{0.037}
\newcommand{\sdBAforties}{0.030}
\newcommand{\sdBAtens}{0.029}
\newcommand{\meanHnineteen}{0.005}
\newcommand{\meanHtens}{0.045}
\newcommand{\rchNineteen}{-0.04}
\newcommand{\rchTwenties}{-0.40}
\newcommand{\rchTens}{-0.58}

\newcommand{\hShareNineteen}{1\%}
\newcommand{\hShareTens}{25\%}

\newcommand{\thShareTens}{40\%}

\newcommand{\gapMed}{0.9\%}
\newcommand{\gapMax}{2.4\%}
\newcommand{\tvhRatio}{33}
\newcommand{\tvbRatio}{0.64}
\newcommand{\cfShareHsixties}{76\%}
\newcommand{\cfShareHtens}{80\%}

\newcommand{\talTens}{0.0059}
\newcommand{\talNineteen}{0.0131}
\newcommand{\talCF}{0.0132}
\newcommand{\covLogFull}{-0.0033}
\newcommand{\covLogExact}{-0.0033}
\newcommand{\covLogNoH}{-0.0007}
\newcommand{\talCFnoH}{0.0112}
\newcommand{\hrChannelShare}{72\%}
\newcommand{\rcfLo}{-0.50}
\newcommand{\rcfHi}{-0.29}
\newcommand{\vindep}{0.0167}
\newcommand{\covShrink}{39\%}

\newcommand{\tvlcRatio}{4.6}
\newcommand{\tvlfRatio}{0.50}
\newcommand{\rtcfNineteen}{0.09}
\newcommand{\rtcfTens}{-0.53}
\newcommand{\nvlfNineteen}{36\%}
\newcommand{\nvlfTens}{51\%}
\newcommand{\sdDrop}{19\%}
\newcommand{\rcfMinPost}{-0.38}
\newcommand{\rcfMaxPost}{-0.31}

\newcommand{\BA}{\mathrm{BA}}
\newcommand{\Var}{\mathrm{Var}}
\newcommand{\Cov}{\mathrm{Cov}}
\newcommand{\logit}{\mathrm{logit}}

\title{Contact, Power, and Balls in Play:\\
A Three-Rate Decomposition of Batting Average\\
and the Disappearance of the .400 Hitter}
\author{Jim Albert\\[2pt]
\small Department of Mathematics and Statistics, Bowling Green State University}
\date{}

\begin{document}
\maketitle

\begin{abstract}
\noindent
Batting average factors exactly as $\BA = c\,\{h + (1-h)\,b\}$, where $c$ is the rate of avoiding a strikeout, $h$ is the home-run rate among non-strikeout at-bats, and $b$ is batting average on balls in play (BABIP). The bracketed term is the rate $f$ at which non-strikeout at-bats become hits, so the decomposition nests the two-factor identity $\BA = c f$. Using the \nQual{} major-league hitters with at least \minAB{} at-bats in 2025, every season from 2004 to 2025, and \nHist{} qualifying seasons from 1901 to 2019, we ask which rates behave like skills and how the corresponding abilities are related. Over \nPairs{} consecutive-season pairs the median year-to-year correlation is \medC{} for $c$, \medH{} for $h$ and \medB{} for $b$; batting average (\medBA{}) is no more repeatable than BABIP. A trivariate logistic-normal random-effects model gives single-season reliabilities of \relC{}, \relH{} and \relB{}, and a talent correlation between contact and power of \rhoCH{} (95\% profile interval \rhoCHlo{} to \rhoCHhi{}), while BABIP talent is only weakly related to either. Historically, the negative covariance between $c$ and $f$, which was near zero in the dead-ball era, explains why batting averages have become less dispersed, and so why no one has hit .400 since 1941. That covariance runs almost entirely through home runs: removing the contact--power channel from the 2010s covariance recovers \hrChannelShare{} of the gap between modern and dead-ball dispersion in batting-average talent.
\end{abstract}


\section{Introduction}\label{sec:intro}

Batting average is the oldest summary of hitting and one of the most studied quantities in sports statistics. It served as the running example in the empirical Bayes work of \citet{EfronMorris1975,EfronMorris1977}, and \citet{Brown2008} used it to compare empirical Bayes and Bayes predictors within a season. These analyses treat a hitter's batting average as a single binomial proportion whose true value is estimated by shrinking the observed average toward a common mean \citep{Morris1983}. Batting average is notoriously unstable from season to season, and a large literature, popular and academic, attributes much of that instability to chance \citep{AlbertBennett2001}.

Batting average is, however, a composite. An at-bat ends in a strikeout or it does not; a non-strikeout at-bat is a home run or a ball in play; and a ball in play is a hit or an out. These three events depend on the hitter in different ways. Avoiding strikeouts reflects pitch recognition and bat-to-ball skill. Home runs need no help from the defense and depend mostly on how hard and at what angle the ball is struck. Whether a ball in play falls for a hit depends also on fielders, parks and chance. \citet{McCracken2001} argued that pitchers have little control over hits on balls in play, and the same reasoning underlies the use of batting average on balls in play (BABIP) as an indicator of hitter luck \citep{Tango2007}.

Several authors have used a partial version of this structure. \citet{Bickel2004} wrote batting average as the product of one minus the strikeout rate and the in-play average to explain why .400 has become so hard to reach. \citet{Albert2005} fitted beta-binomial random-effects models to several hitting rates and found that strikeout and home-run rates are far more ability-driven than batting average, and \citet{Albert2016} showed that predicting component rates separately and recombining them improves forecasts of aggregate measures. What has not been examined is the joint structure of the components: how the underlying abilities are correlated once sampling noise is removed, and how that correlation has changed over the history of the game.

This paper writes batting average exactly as the product of three rates and asks three questions:
\begin{enumerate}
\item Which rates behave like stable skills, and how much of each rate's cross-sectional variation reflects ability rather than chance (Sections~\ref{sec:repeat} and~\ref{sec:model})?
\item How are the three abilities related once chance is removed, and what does the dependence imply for estimating an individual hitter's ability (Section~\ref{sec:model})?
\item \citet{Gould1986,Gould1996} attributed the disappearance of the .400 hitter to a narrowing of the distribution of batting averages. Which components drove that narrowing (Section~\ref{sec:history})?
\end{enumerate}
The answers are that strikeout avoidance and home-run hitting are skills while BABIP is mostly noise; that contact and power abilities are strongly negatively correlated; and that the emergence of this contact--power trade-off, rather than a convergence of hitting talent, accounts for most of the reduced dispersion of modern batting averages.

\section{Data and the decomposition}\label{sec:data}

\subsection{The identity}

For a hitter with $AB$ at-bats, $SO$ strikeouts, $HR$ home runs and $H$ hits, define
\begin{equation}\label{eq:rates}
c = \frac{AB - SO}{AB}, \qquad h = \frac{HR}{AB - SO}, \qquad b = \frac{H - HR}{AB - SO - HR}.
\end{equation}
We call $c$ the contact rate, $h$ the home-run rate and $b$ BABIP. Our $b$ differs from the conventional BABIP only in omitting sacrifice flies from the denominator. Because $ch = HR/AB$ and $c(1-h)b = (H-HR)/AB$,
\begin{equation}\label{eq:identity}
\BA = \frac{H}{AB} = c\,\{h + (1-h)\,b\} = c\,f, \qquad f = \frac{H}{AB-SO} = h + (1-h)\,b,
\end{equation}
exactly, for every hitter and season. The finishing rate $f$ is the proportion of non-strikeout at-bats that become hits. Writing $\BA = cf$ gives the coarser two-factor view used by \citet{Bickel2004}; splitting $f$ into $h$ and $b$ separates the part of finishing that the defense cannot affect from the part it can. Among the 2025 hitters the identity holds to within \identityErr{}; \nZeroHR{} of them hit no home runs.

As an illustration, Aaron Judge had 541 at-bats in 2025 and struck out 160 times, so $c = 381/541 = \judgeC{}$. Of the 381 non-strikeout at-bats, 53 were home runs ($h = \judgeH{}$), and 126 of the remaining 328 balls in play fell for hits ($b = \judgeB{}$). His 179 hits in 541 at-bats give $\BA = \judgeBA{}$.

\subsection{Samples}

All data come from the Lahman database \citep{Lahman2026}, analyzed in R \citep{R2024}. Stints with different teams in a season are combined. The main sample consists of the \nQual{} hitters with at least \minAB{} at-bats in 2025. For repeatability we use every season from 2004 to 2025 with the same threshold. For the historical analysis we use the \nHist{} American, National and Federal League player-seasons from 1901 to 2019 with at least \minAB{} at-bats and a recorded strikeout total; the Negro Leagues, recently added to the database, are excluded because their strikeout records are incomplete. Seasons are grouped by decade.

\section{The cross-section in 2025}\label{sec:cross}

\subsection{Equal averages, different compositions}

Table~\ref{tab:cases} shows four 2025 hitters whose averages fall between .311 and .331. Judge and Jacob Wilson reached similar averages by opposite routes: Judge with a low contact rate, a home-run rate nearly five times Wilson's and a high BABIP; Wilson with the second-highest contact rate among qualifiers and little power. \powerName{} had the highest home-run rate, and \babipName{} the highest BABIP, among the other hitters within twenty points of Judge's average.

\begin{table}[t]
\centering\small
\caption{Four 2025 hitters with similar batting averages and different combinations of contact rate $c$, home-run rate $h$ and BABIP $b$.}
\label{tab:cases}
\begin{tabular}{lrrrrrrrr}
\toprule
Player & $AB$ & $H$ & $SO$ & $HR$ & BA & $c$ & $h$ & $b$ \\
\midrule
Aaron Judge & 541 & 179 & 160 & 53 & 0.331 & 0.704 & 0.139 & 0.384 \\
Miguel Andujar & 321 & 102 &  49 & 10 & 0.318 & 0.847 & 0.037 & 0.351 \\
Jonathan Aranda & 370 & 117 & 107 & 14 & 0.316 & 0.711 & 0.053 & 0.414 \\
Jacob Wilson & 486 & 151 &  39 & 13 & 0.311 & 0.920 & 0.029 & 0.318
 \\
\bottomrule
\end{tabular}
\end{table}

\begin{figure}[t]
\centering
\includegraphics[width=\textwidth]{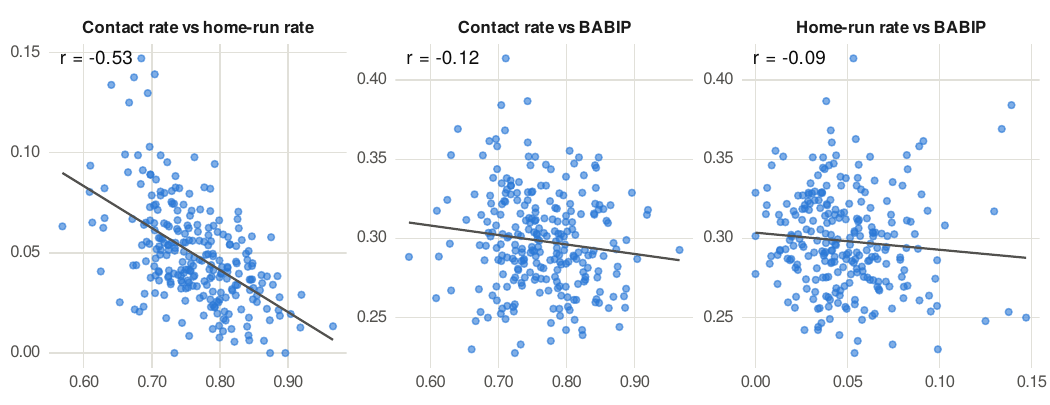}
\caption{Pairwise relationships among contact rate $c$, home-run rate $h$ and BABIP $b$ for the \nQual{} hitters with at least \minAB{} at-bats in 2025. Lines are least-squares fits; $r$ is the Pearson correlation.}
\label{fig:pairs}
\end{figure}

Figure~\ref{fig:pairs} shows the pairwise relationships among the three rates for all \nQual{} hitters. Contact and home-run rates are strongly negatively correlated ($r = \rch{}$, 95\% interval \rchLo{} to \rchHi{}). BABIP is only weakly related to contact ($r = \rcb{}$) and to home-run rate ($r = \rhb{}$). At the two-factor level the correlation between $c$ and $f$ is \rcf{} (\rcfLo{} to \rcfHi{}). The familiar observation that hitters who rarely strike out do less damage when they make contact is therefore, in 2025, primarily a statement about home runs.

\subsection{A variance decomposition}\label{sec:vardecomp}

Taking logarithms of \eqref{eq:identity} gives $\log \BA = \log c + \log f$, so the cross-sectional variance decomposes exactly as
\begin{equation}\label{eq:vlog}
\Var(\log \BA) = \Var(\log c) + \Var(\log f) + 2\,\Cov(\log c, \log f).
\end{equation}
In 2025 the three terms are \vc{}, \vf{} and \twocov{}, summing to \vba{}. The two factors contribute comparable variance, but because they are negatively correlated the observed dispersion of batting average is \covShrink{} smaller than it would be if $c$ and $f$ varied independently (\vindep{}). This covariance term is the quantity that drives the historical argument of Section~\ref{sec:history}.

Because $f$ is a sum rather than a product, $\Var(\log f)$ cannot be split exactly. Linearizing about the means $\bar h$, $\bar b$ and $\bar f$,
\begin{equation}\label{eq:delta}
\Var(\log f) \approx \frac{1}{\bar f^2}\Big\{(1-\bar b)^2\,\Var(h) + (1-\bar h)^2\,\Var(b) + 2(1-\bar b)(1-\bar h)\,\Cov(h,b)\Big\}.
\end{equation}
In 2025 the three terms are \aHR{}, \aB{} and \aCov{}, shares of \aHRshare{}, \aBshare{} and \aCovshare{}; their sum is within \aGap{} of the exact $\Var(\log f)$. On observed rates, then, BABIP accounts for most of the variation in finishing. Much of that variation is sampling noise, however. Subtracting the average binomial variance $\overline{p(1-p)/n}$ from each observed variance removes \bNoiseCur{} of the variance of $b$ but only \hNoiseCur{} of that of $h$, after which the home-run and BABIP terms account for \taHRshare{} and \taBshare{} of the total. Once chance is removed, power and BABIP contribute about equally to differences in finishing ability.

\section{Which rates are skills?}\label{sec:repeat}

A rate that measures ability should be repeatable. For each pair of consecutive seasons from 2004 to 2025 we correlate each rate across the hitters who qualified in both seasons. The 2020 season, in which almost no one reached \minAB{} at-bats, drops out, leaving \nPairs{} pairs with a median of \medPairN{} hitters each.

\begin{figure}[t]
\centering
\includegraphics[width=\textwidth]{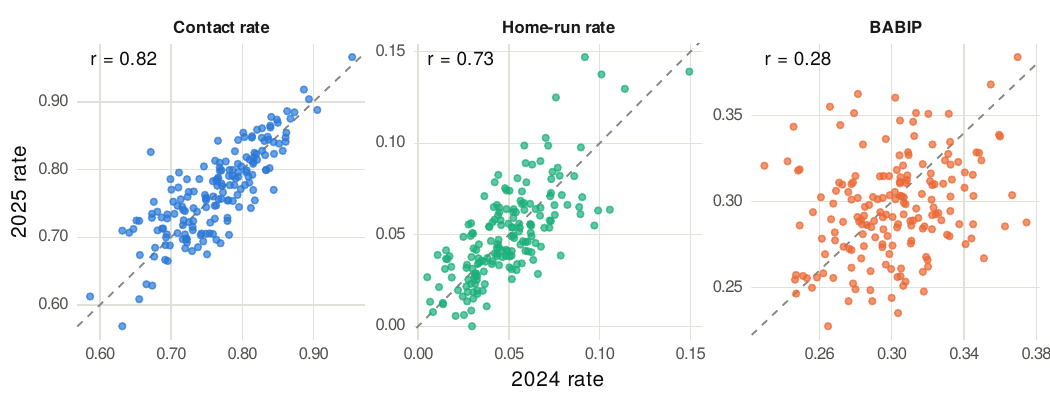}
\caption{2024 and 2025 rates for the \nBoth{} hitters with at least \minAB{} at-bats in both seasons. The dashed line is $y = x$.}
\label{fig:repeat}
\end{figure}

\begin{figure}[t]
\centering
\includegraphics[width=0.72\textwidth]{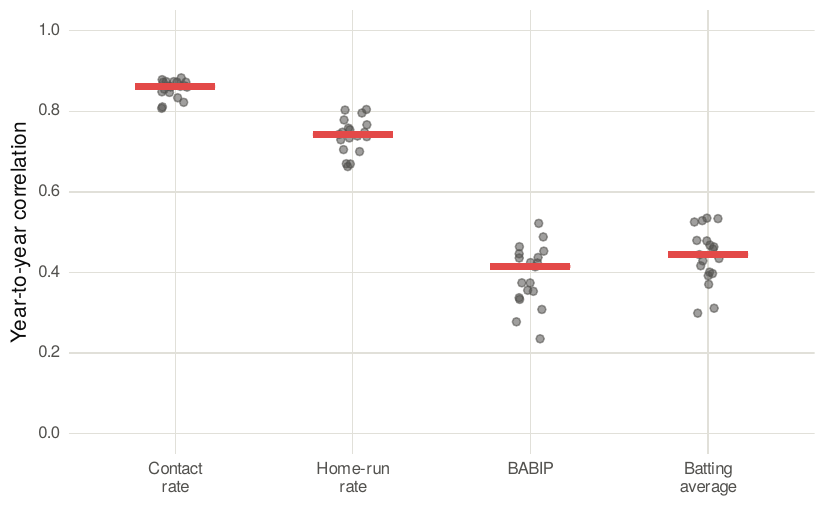}
\caption{Year-to-year correlations for each of the \nPairs{} consecutive-season pairs from 2004 to 2025. Horizontal bars mark the medians.}
\label{fig:stability}
\end{figure}

Figure~\ref{fig:repeat} shows the 2024--2025 pair (\nBoth{} hitters). Contact and home-run rates repeat closely ($r = \rCtwo{}$ and \rHtwo{}); BABIP barely does ($r = \rBtwo{}$). Figure~\ref{fig:stability} shows that the pattern holds in every pair. The year-to-year correlation ranges from \minC{} to \maxC{} for $c$ (median \medC{}), from \minH{} to \maxH{} for $h$ (median \medH{}), and from \minB{} to \maxB{} for $b$ (median \medB{}). For batting average it ranges from \minBA{} to \maxBA{} (median \medBA{}): the composite is no more repeatable than its noisiest component.

\section{A trivariate random-effects model}\label{sec:model}

\subsection{Model and estimation}

Repeatability uses two seasons. A random-effects model extracts the same information from a single season and also estimates how the three abilities are related. Because the three outcomes are nested, each can be modeled as binomial conditional on the one before it. For hitter $i$,
\begin{align}
AB_i - SO_i &\sim \mathrm{Bin}(AB_i,\, c_i), & \logit(c_i) &= \mu_c + u_i, \notag\\
HR_i &\sim \mathrm{Bin}(AB_i - SO_i,\, h_i), & \logit(h_i) &= \mu_h + v_i, \label{eq:model}\\
H_i - HR_i &\sim \mathrm{Bin}(AB_i - SO_i - HR_i,\, b_i), & \logit(b_i) &= \mu_b + w_i, \notag
\end{align}
with $(u_i, v_i, w_i)^\top \sim \mathrm{N}_3(\mathbf 0, \Sigma)$ and $\Sigma$ unstructured, with standard deviations $\sigma_c, \sigma_h, \sigma_b$ and correlations $\rho_{ch}, \rho_{cb}, \rho_{hb}$. Given the abilities, the three binomial outcomes are conditionally independent because they are determined by disjoint sets of at-bats. We fit \eqref{eq:model} by maximum likelihood with the Laplace approximation in \texttt{lme4} \citep{Bates2015}, stacking the three outcomes and specifying an unstructured covariance for the three random intercepts, and compute profile-likelihood intervals for the variance components.

Table~\ref{tab:params} gives the estimates for 2025. The means correspond to league rates of \pmuC{} for contact, \pmuH{} for home runs and \pmuB{} for BABIP. On the logit scale the spread of talent is largest for home runs ($\sigma_h = \sigH{}$), intermediate for contact ($\sigma_c = \sigC{}$), and small for BABIP ($\sigma_b = \sigB{}$).

\begin{table}[t]
\centering\small
\caption{Maximum likelihood estimates for model \eqref{eq:model}, 2025 season (logit scale). Wald intervals for the means; profile-likelihood intervals for the standard deviations and correlations.}
\label{tab:params}
\begin{tabular}{llr@{\hspace{1.5em}}l}
\toprule
Parameter & Interpretation & Estimate & 95\% interval \\
\midrule
$\mu_c$ & league contact rate (\pmuC{}) & \muC{} & (\muClo{}, \muChi{}) \\
$\mu_h$ & league home-run rate (\pmuH{}) & \muH{} & (\muHlo{}, \muHhi{}) \\
$\mu_b$ & league BABIP (\pmuB{}) & \muB{} & (\muBlo{}, \muBhi{}) \\
$\sigma_c$ & talent SD, contact & \sigC{} & (\sigClo{}, \sigChi{}) \\
$\sigma_h$ & talent SD, home runs & \sigH{} & (\sigHlo{}, \sigHhi{}) \\
$\sigma_b$ & talent SD, BABIP & \sigB{} & (\sigBlo{}, \sigBhi{}) \\
$\rho_{ch}$ & talent correlation, contact and home runs & \rhoCH{} & (\rhoCHlo{}, \rhoCHhi{}) \\
$\rho_{cb}$ & talent correlation, contact and BABIP & \rhoCB{} & (\rhoCBlo{}, \rhoCBhi{}) \\
$\rho_{hb}$ & talent correlation, home runs and BABIP & \rhoHB{} & (\rhoHBlo{}, \rhoHBhi{}) \\
\bottomrule
\end{tabular}
\end{table}

\subsection{Talent correlations}

The estimated correlation between contact and home-run abilities is $\hat\rho_{ch} = \rhoCH{}$, noticeably stronger than the raw correlation of \rch{}. Sampling noise inflates the variance of each observed rate without changing their covariance, so observed correlations are attenuated toward zero \citep{Spearman1904}; the model removes that attenuation. Covariances between observed rates need no such correction, because the three rates are computed from disjoint sets of at-bats and their sampling errors are independent. BABIP ability is only weakly related to the other two abilities ($\hat\rho_{cb} = \rhoCB{}$, $\hat\rho_{hb} = \rhoHB{}$), and neither interval clearly excludes zero.

\subsection{Single-season reliability}

By the delta method, the sampling variance of hitter $i$'s observed logit for a rate $p$ based on $n$ trials is approximately $\tau_i^2 = 1/\{n\,\hat p(1-\hat p)\}$; we add $\tfrac12$ to success and failure counts so that a season with no home runs remains finite. The ratio $w_i = \sigma^2/(\sigma^2 + \tau_i^2)$ is the fraction of between-hitter variation in that rate that reflects ability, the reliability of classical test theory \citep{LordNovick1968}. Averaged over the 2025 hitters the reliability is \relC{} for contact, \relH{} for home runs and \relB{} for BABIP. The mean noise standard deviations are \tauC{}, \tauH{} and \tauB{}: home-run rates are noisier than BABIP in absolute terms, but their talent spread is more than six times larger. These single-season reliabilities agree closely with the multi-season correlations of Section~\ref{sec:repeat}, as they should, since reliability and test--retest correlation estimate the same quantity when ability is stable.

\subsection{Trivariate shrinkage}

Let $z_i$ hold hitter $i$'s observed logits minus $(\mu_c, \mu_h, \mu_b)$ and let $T_i = \mathrm{diag}(\tau^2_{i,c}, \tau^2_{i,h}, \tau^2_{i,b})$. Under a normal approximation to the binomial likelihood on the logit scale, the posterior mean of his abilities is
\begin{equation}\label{eq:shrink}
\hat\theta_i = \Sigma\,(\Sigma + T_i)^{-1} z_i.
\end{equation}
When $\Sigma$ is diagonal, \eqref{eq:shrink} reduces to shrinking each rate separately by its own reliability, the familiar James--Stein form \citep{EfronMorris1975}. When it is not, the off-diagonal terms let each observed rate inform the estimates of the other abilities.

\begin{figure}[t]
\centering
\includegraphics[width=0.82\textwidth]{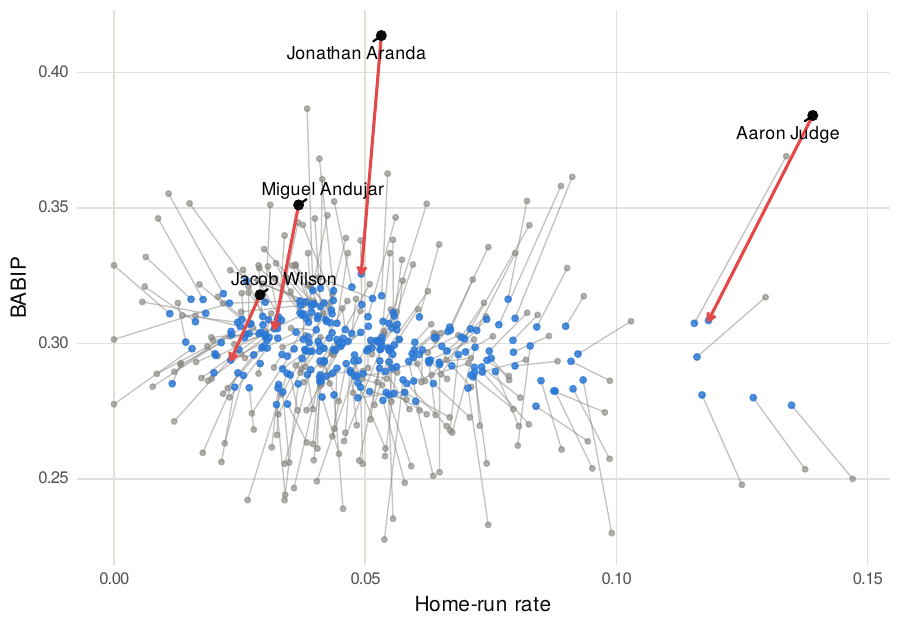}
\caption{Observed 2025 home-run rate and BABIP (gray) and model estimates (blue) for the \nQual{} qualifiers. Arrows run from observed to estimated; the four hitters of Table~\ref{tab:cases} are highlighted. The arrows are nearly vertical: BABIP is shrunk strongly toward the league mean, home-run rate much less.}
\label{fig:shrink}
\end{figure}

Table~\ref{tab:shrink} and Figure~\ref{fig:shrink} show the conditional modes from the fitted model, which use the full binomial likelihood. Contact and home-run rates barely move; the mean absolute adjustments are \mnAbsDC{} and \mnAbsDH{}. BABIP moves much more (mean adjustment \mnAbsDB{}): Judge's observed .384 becomes an estimated ability of \judgeBs{}, close to the league mean of \pmuB{}, and Aranda's .414 becomes \babipShr{}. The correlations matter for individual hitters. Shrinking Judge's BABIP by its own reliability alone gives \judgeBuni{}, whereas \eqref{eq:shrink} gives \judgeBmulti{}, because his high home-run rate, through $\hat\rho_{hb} < 0$, points to a lower BABIP ability. Averaged over hitters, the correlations shift the BABIP estimate by \mnAbsBorrow{}.

\begin{table}[t]
\centering\small
\caption{Observed 2025 rates and model estimates (conditional modes) for the four hitters of Table~\ref{tab:cases}.}
\label{tab:shrink}
\begin{tabular}{lrrrrrr}
\toprule
 & \multicolumn{2}{c}{Contact rate} & \multicolumn{2}{c}{Home-run rate} & \multicolumn{2}{c}{BABIP} \\
\cmidrule(lr){2-3}\cmidrule(lr){4-5}\cmidrule(lr){6-7}
Player & observed & estimate & observed & estimate & observed & estimate \\
\midrule
Aaron Judge & 0.704 & 0.694 & 0.139 & 0.118 & 0.384 & 0.308 \\
Miguel Andujar & 0.847 & 0.833 & 0.037 & 0.032 & 0.351 & 0.305 \\
Jonathan Aranda & 0.711 & 0.712 & 0.053 & 0.049 & 0.414 & 0.326 \\
Jacob Wilson & 0.920 & 0.899 & 0.029 & 0.023 & 0.318 & 0.294
 \\
\bottomrule
\end{tabular}
\end{table}

\section{A historical perspective on the .400 hitter}\label{sec:history}

\subsection{Gould's argument and the two-factor decomposition}

\begin{figure}[t]
\centering
\includegraphics[width=\textwidth]{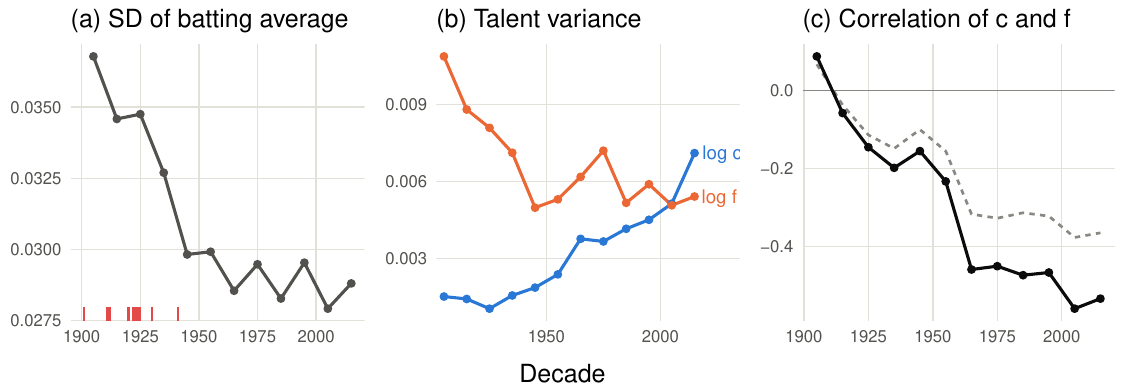}
\caption{Qualifying AL, NL and FL hitters by decade, 1900s--2010s. (a) Standard deviation of batting average; ticks mark the \nFourHundred{} seasons of .400 or better. (b) Talent-scale variances of $\log c$ and $\log f$ (observed variances minus average sampling variance). (c) Correlation between $c$ and $f$, observed (dashed) and with sampling noise removed (solid).}
\label{fig:history}
\end{figure}

Among American, National and Federal League hitters with at least \minAB{} at-bats there have been \nFourHundred{} seasons of .400 or better since 1901, the last in \lastFourHundred{}. \citet{Gould1986,Gould1996} argued that .400 disappeared not because hitters became worse but because the distribution of batting averages narrowed: as play improved and the talent pool deepened, the distance between ordinary and exceptional hitters shrank, trimming both tails. The standard deviation of batting average among qualifiers did fall, from \sdBAnineteen{} in the 1900s to \sdBAforties{} in the 1940s (about \sdDrop{}), and has been nearly constant since (\sdBAtens{} in the 2010s; Figure~\ref{fig:history}a).

Decomposition \eqref{eq:vlog} shows where the narrowing came from. Table~\ref{tab:logdecomp} reports its terms by decade, on the observed scale and, after subtracting the average delta-method sampling variance $(1-p)/(np)$ from $\Var(\log c)$ and $\Var(\log f)$, on the talent scale. Sampling noise matters most for $f$, accounting for \nvlfNineteen{} of the observed variance of $\log f$ in the 1900s and \nvlfTens{} in the 2010s. Three trends emerge (Figure~\ref{fig:history}b,c).
\begin{enumerate}
\item Strikeout avoidance did not homogenize. Its talent variance on the log scale grew by a factor of about \tvlcRatio{} from the 1900s to the 2010s.
\item The talent variance of finishing roughly halved (ratio \tvlfRatio{}), consistent with Gould's picture of a converging talent pool.
\item The two factors were essentially uncorrelated in the dead-ball era (talent correlation \rtcfNineteen{}), while since the 1960s the observed correlation has stayed between \rcfMinPost{} and \rcfMaxPost{} (talent correlation \rtcfTens{} in the 2010s).
\end{enumerate}

\begin{table}[t]
\centering\footnotesize
\setlength{\tabcolsep}{4pt}
\caption{Decomposition \eqref{eq:vlog} by decade for qualifying hitters. Columns 3--8 are observed quantities; the last three remove binomial sampling noise from the variances of $\log c$ and $\log f$ and from their correlation.}
\label{tab:logdecomp}
\begin{tabular}{lrrrrrrrrrr}
\toprule
 & & & & & & & & \multicolumn{3}{c}{Talent scale} \\
\cmidrule(lr){9-11}
Decade & $n$ & SD(BA) & $\Var(\log c)$ & $\Var(\log f)$ & $2\Cov$ & $\Var(\log \BA)$ & $r(c,f)$ & $\Var(\log c)$ & $\Var(\log f)$ & $r(c,f)$ \\
\midrule
1900s & 1105 & 0.037 & 0.0018 & 0.0170 & 0.0007 & 0.0195 & 0.07 & 0.0015 & 0.0109 & 0.09 \\
1910s & 1276 & 0.035 & 0.0017 & 0.0147 & -0.0004 & 0.0160 & -0.04 & 0.0014 & 0.0088 & -0.06 \\
1920s & 1241 & 0.035 & 0.0012 & 0.0132 & -0.0009 & 0.0136 & -0.11 & 0.0011 & 0.0081 & -0.15 \\
1930s & 1296 & 0.033 & 0.0018 & 0.0123 & -0.0013 & 0.0128 & -0.15 & 0.0016 & 0.0071 & -0.20 \\
1940s & 1211 & 0.030 & 0.0021 & 0.0107 & -0.0010 & 0.0119 & -0.10 & 0.0019 & 0.0050 & -0.16 \\
1950s & 1227 & 0.030 & 0.0027 & 0.0111 & -0.0017 & 0.0121 & -0.16 & 0.0024 & 0.0053 & -0.23 \\
1960s & 1581 & 0.029 & 0.0042 & 0.0121 & -0.0044 & 0.0118 & -0.32 & 0.0038 & 0.0062 & -0.46 \\
1970s & 2038 & 0.029 & 0.0040 & 0.0130 & -0.0046 & 0.0124 & -0.33 & 0.0037 & 0.0072 & -0.45 \\
1980s & 2114 & 0.028 & 0.0046 & 0.0110 & -0.0044 & 0.0112 & -0.31 & 0.0042 & 0.0052 & -0.47 \\
1990s & 2229 & 0.030 & 0.0050 & 0.0115 & -0.0048 & 0.0116 & -0.32 & 0.0045 & 0.0059 & -0.47 \\
2000s & 2545 & 0.028 & 0.0056 & 0.0105 & -0.0057 & 0.0104 & -0.38 & 0.0051 & 0.0051 & -0.56 \\
2010s & 2494 & 0.029 & 0.0077 & 0.0111 & -0.0066 & 0.0122 & -0.37 & 0.0071 & 0.0054 & -0.53
 \\
\bottomrule
\end{tabular}
\end{table}

A counterfactual makes the role of the covariance precise. On the talent scale, the variance of $\log \BA$ fell from \talNineteen{} in the 1900s to \talTens{} in the 2010s. Combining the 2010s talent variances of $\log c$ and $\log f$ with the 1900s covariance gives \talCF{}, essentially the dead-ball level. The growth in the spread of contact ability and the narrowing of finishing ability roughly offset, and the reduced spread of modern batting averages is due almost entirely to the stronger negative covariance between the two factors. Gould's account is therefore incomplete: a .400 season requires excellence in both $c$ and $f$ at once, and the modern game rarely produces that combination.

\subsection{The trade-off runs through home runs}

\begin{figure}[t]
\centering
\includegraphics[width=0.9\textwidth]{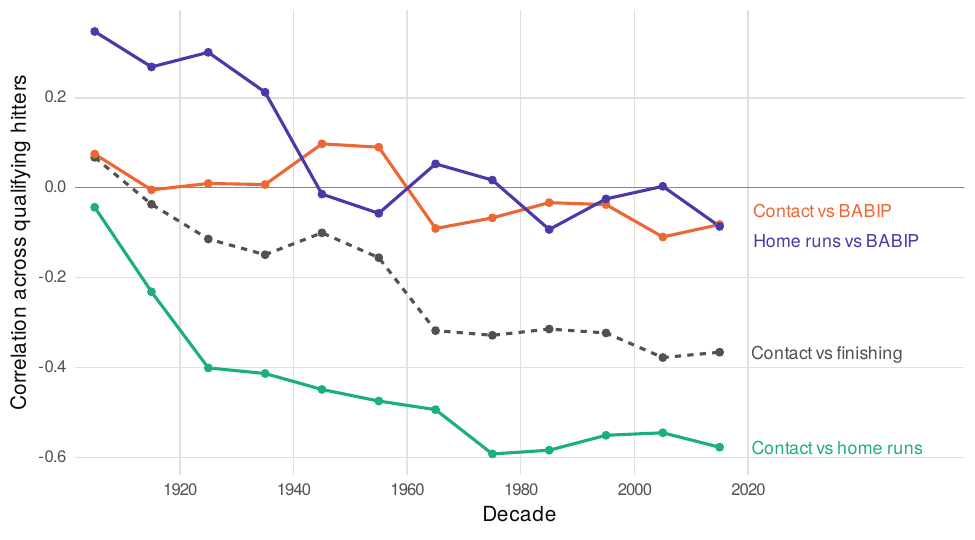}
\caption{Observed correlations among the three rates, and between contact and finishing (dashed), for qualifying hitters in each decade, 1900s--2010s.}
\label{fig:tradeoffs}
\end{figure}

The three-rate decomposition identifies which part of finishing trades off against contact. Figure~\ref{fig:tradeoffs} and Table~\ref{tab:decades} show the correlations among $c$, $h$ and $b$ by decade. The correlation between contact and home runs was close to zero in the 1900s (\rchNineteen{}), when home runs were rare, fell to \rchTwenties{} in the 1920s as the home run entered the game, and reached \rchTens{} by the 2010s. The correlation between contact and BABIP was small in every decade. Hitters who struck out more did not do systematically better or worse on balls in play; they hit more home runs.

Linearizing $f$ makes the attribution explicit:
\begin{equation}\label{eq:covsplit}
\Cov(c, f) \approx (1-\bar b)\,\Cov(c, h) + (1-\bar h)\,\Cov(c, b).
\end{equation}
The home-run term accounts for \cfShareHsixties{} of $\Cov(c,f)$ in the 1960s and \cfShareHtens{} in the 2010s, and for more than all of it from the 1920s through the 1950s, when the BABIP term was slightly positive.

\begin{table}[t]
\centering\footnotesize
\setlength{\tabcolsep}{4pt}
\caption{Correlations among the three rates by decade for qualifying hitters. $\bar h$ is the mean home-run rate. Columns headed $r$ are observed correlations; ``talent $r$'' columns divide the observed covariance by noise-corrected standard deviations. The last column is the home-run term's share of approximation \eqref{eq:delta}.}
\label{tab:decades}
\begin{tabular}{lrrrrrrrrrr}
\toprule
Decade & $n$ & SD(BA) & $\bar h$ & $r(c,f)$ & $r(c,h)$ & $r(c,b)$ & $r(h,b)$ & talent $r(c,h)$ & talent $r(c,b)$ & HR share \\
\midrule
1900s & 1105 & 0.037 & 0.005 & 0.07 & -0.04 & 0.08 & 0.35 & -0.06 & 0.10 & 1\% \\
1920s & 1241 & 0.035 & 0.014 & -0.11 & -0.40 & 0.01 & 0.30 & -0.46 & 0.01 & 10\% \\
1940s & 1211 & 0.030 & 0.020 & -0.10 & -0.45 & 0.10 & -0.01 & -0.51 & 0.16 & 18\% \\
1960s & 1581 & 0.029 & 0.033 & -0.32 & -0.49 & -0.09 & 0.05 & -0.56 & -0.15 & 24\% \\
1980s & 2114 & 0.028 & 0.031 & -0.31 & -0.58 & -0.03 & -0.09 & -0.66 & -0.05 & 24\% \\
2000s & 2545 & 0.028 & 0.042 & -0.38 & -0.54 & -0.11 & 0.00 & -0.62 & -0.19 & 27\% \\
2010s & 2494 & 0.029 & 0.045 & -0.37 & -0.58 & -0.08 & -0.09 & -0.67 & -0.13 & 25\%
 \\
\bottomrule
\end{tabular}
\end{table}

\subsection{Home runs became part of finishing}

\begin{figure}[t]
\centering
\includegraphics[width=0.9\textwidth]{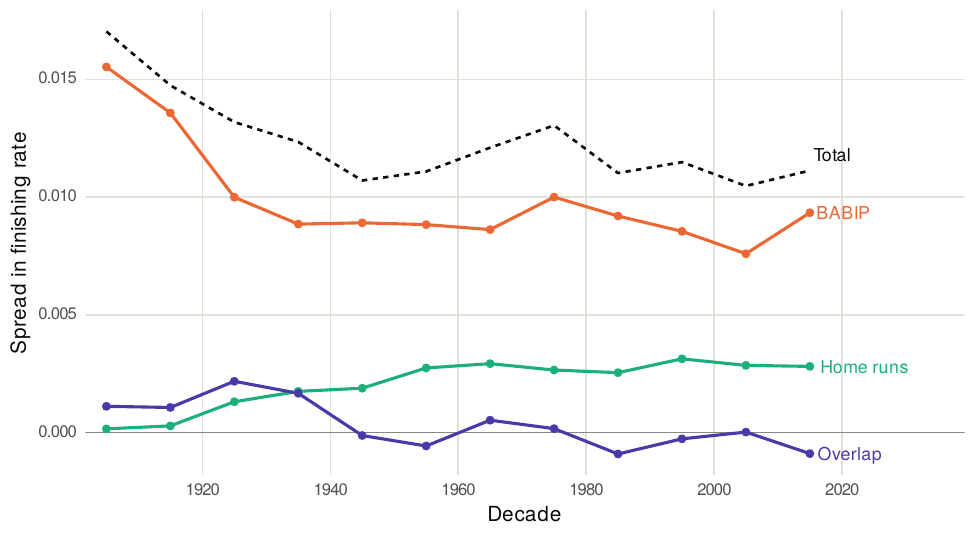}
\caption{Approximation \eqref{eq:delta} of $\Var(\log f)$ by decade: the home-run term, the BABIP term and the covariance (overlap) term. The dashed line is the exact $\Var(\log f)$.}
\label{fig:fdecomp}
\end{figure}

Figure~\ref{fig:fdecomp} applies approximation \eqref{eq:delta} by decade. In the 1900s the average qualifying hitter homered in about one of every 200 non-strikeout at-bats ($\bar h = \meanHnineteen{}$), and home runs accounted for only \hShareNineteen{} of the variation in finishing rates; finishing was essentially BABIP. By the 2010s the mean home-run rate was about nine times higher ($\bar h = \meanHtens{}$), and home runs accounted for \hShareTens{} of that variation (\thShareTens{} once sampling noise is removed). Over the century the talent variance of $h$ grew about \tvhRatio{}-fold, while that of $b$ fell to \tvbRatio{} of its 1900s level. Power became a second, growing dimension along which hitters differ, and it is the dimension that trades off against contact.

\subsection{How much did the contact--power trade-off matter?}

Equation~\eqref{eq:covsplit} supports a narrower counterfactual than the one above: keep everything about the 2010s except the contact--home-run channel. Dividing \eqref{eq:covsplit} by $\bar c \bar f$ approximates $\Cov(\log c, \log f)$; for the 2010s it gives \covLogFull{}, matching the exact value of \covLogExact{}. Setting the home-run term to zero leaves \covLogNoH{}, and the talent-scale variance of $\log \BA$ becomes \talCFnoH{}. That is \hrChannelShare{} of the way from the actual 2010s value (\talTens{}) to the value with dead-ball covariance (\talCF{}). Roughly three-quarters of the reduced dispersion of modern batting-average talent is attributable to the trade-off between avoiding strikeouts and hitting home runs.

\section{Discussion}\label{sec:discussion}

The identity $\BA = c\{h + (1-h)b\}$ separates batting average into two components that behave like skills and one that behaves largely like noise. Strikeout avoidance and home-run rate are highly repeatable and highly reliable within a single season; BABIP is neither, and it is the main reason batting average is so unstable. Because the identity is exact, the variance decompositions and counterfactuals of Sections~\ref{sec:cross} and~\ref{sec:history} require no modeling assumptions beyond the linearization in \eqref{eq:delta} and \eqref{eq:covsplit}, whose accuracy we check in the Appendix. The random-effects model adds estimates of the talent correlations and a trivariate shrinkage estimator that lets each rate inform the others, extending the component approach of \citet{Albert2016} by modeling the dependence between components.

The historical analysis refines Gould's explanation for the disappearance of the .400 hitter. Batting averages did become less dispersed, but not because hitters became more alike: the spread of contact ability grew substantially. What changed is that, beginning in the 1920s, the hitters who make the most contact have tended to hit the fewest home runs. A modern .400 season would require a hitter who rarely strikes out, hits for power, and enjoys a favorable season on balls in play, a combination that has been rare since the home run became central to the game.

Several limitations should be noted. First, BABIP reflects both contact quality and defense; batted-ball tracking data could separate the two. Second, home-run rates depend on parks and on the baseball itself, so part of what we call power is environmental. Third, at-bats exclude walks, hit-by-pitches and sacrifices, so the rates describe the at-bat outcomes that enter batting average rather than all plate appearances. Fourth, the \minAB{}-at-bat threshold selects hitters whose teams kept playing them, which may truncate the lower tails of the rates. Fifth, the model is fit to one season at a time; a model with player effects shared across seasons, or a fully Bayesian fit with priors on $\Sigma$ \citep{GelmanHill2007}, would give sharper estimates and full posterior uncertainty for the correlations. Finally, the historical analysis establishes when the contact--power trade-off emerged, not why. Changes in pitching, in hitters' approaches, and in how teams value strikeouts relative to power are all candidates.

\section*{Acknowledgments and use of AI tools}

The analyses and this manuscript were produced in collaboration with Claude (Anthropic), a large language model. The author posed the research questions, chose which directions to pursue, and reviewed the results. Claude wrote the R code, carried out the computations, drafted the text, and assembled the references. Claude is not listed as an author because it cannot take responsibility for the work; the author takes full responsibility for the content, including any remaining errors. A single R script reproduces every number, table and figure and is available from the author.

\bibliographystyle{apalike}
\bibliography{refs}

@article{Albert2016,
  author  = {Albert, Jim},
  title   = {Improved Component Predictions of Batting and Pitching Measures},
  journal = {Journal of Quantitative Analysis in Sports},
  year    = {2016},
  volume  = {12},
  number  = {2},
  pages   = {73--85},
  doi     = {10.1515/jqas-2015-0063}
}

@book{AlbertBennett2001,
  author    = {Albert, Jim and Bennett, Jay},
  title     = {Curve Ball: Baseball, Statistics, and the Role of Chance in the Game},
  publisher = {Copernicus Books},
  address   = {New York},
  year      = {2001}
}

@article{Bates2015,
  author  = {Bates, Douglas and M{\"a}chler, Martin and Bolker, Ben and Walker, Steve},
  title   = {Fitting Linear Mixed-Effects Models Using {lme4}},
  journal = {Journal of Statistical Software},
  year    = {2015},
  volume  = {67},
  number  = {1},
  pages   = {1--48},
  doi     = {10.18637/jss.v067.i01}
}

@article{Brown2008,
  author  = {Brown, Lawrence D.},
  title   = {In-Season Prediction of Batting Averages: A Field Test of Empirical {Bayes} and {Bayes} Methodologies},
  journal = {The Annals of Applied Statistics},
  year    = {2008},
  volume  = {2},
  number  = {1},
  pages   = {113--152}
}

@article{EfronMorris1975,
  author  = {Efron, Bradley and Morris, Carl},
  title   = {Data Analysis Using {Stein's} Estimator and Its Generalizations},
  journal = {Journal of the American Statistical Association},
  year    = {1975},
  volume  = {70},
  number  = {350},
  pages   = {311--319}
}

@article{EfronMorris1977,
  author  = {Efron, Bradley and Morris, Carl},
  title   = {{Stein's} Paradox in Statistics},
  journal = {Scientific American},
  year    = {1977},
  volume  = {236},
  number  = {5},
  pages   = {119--127}
}

@manual{Lahman2026,
  author = {Friendly, Michael and Dalzell, Chris and Monkman, Martin and Murphy, Dennis},
  title  = {{Lahman}: Sean `{Lahman}' Baseball Database},
  year   = {2026},
  note   = {R package version 14.0-0},
  url    = {https://CRAN.R-project.org/package=Lahman}
}

@book{GelmanHill2007,
  author    = {Gelman, Andrew and Hill, Jennifer},
  title     = {Data Analysis Using Regression and Multilevel/Hierarchical Models},
  publisher = {Cambridge University Press},
  address   = {Cambridge},
  year      = {2007}
}

@article{Gould1986,
  author  = {Gould, Stephen Jay},
  title   = {Entropic Homogeneity Isn't Why No One Hits .400 Any More},
  journal = {Discover},
  year    = {1986},
  month   = aug
}

@book{Gould1996,
  author    = {Gould, Stephen Jay},
  title     = {Full House: The Spread of Excellence from {Plato} to {Darwin}},
  publisher = {Harmony Books},
  address   = {New York},
  year      = {1996}
}

@book{LordNovick1968,
  author    = {Lord, Frederic M. and Novick, Melvin R.},
  title     = {Statistical Theories of Mental Test Scores},
  publisher = {Addison-Wesley},
  address   = {Reading, MA},
  year      = {1968}
}

@misc{McCracken2001,
  author       = {McCracken, Voros},
  title        = {Pitching and Defense: How Much Control Do Hurlers Have?},
  howpublished = {Baseball Prospectus},
  year         = {2001},
  month        = jan,
  note         = {January 23, 2001},
  url          = {https://www.baseballprospectus.com/news/article/878/pitching-and-defense-how-much-control-do-hurlers-have/}
}

@article{Morris1983,
  author  = {Morris, Carl N.},
  title   = {Parametric Empirical {Bayes} Inference: Theory and Applications},
  journal = {Journal of the American Statistical Association},
  year    = {1983},
  volume  = {78},
  number  = {381},
  pages   = {47--55}
}

@manual{R2024,
  author       = {{R Core Team}},
  title        = {R: A Language and Environment for Statistical Computing},
  organization = {R Foundation for Statistical Computing},
  address      = {Vienna, Austria},
  year         = {2024},
  url          = {https://www.R-project.org/}
}

@article{Spearman1904,
  author  = {Spearman, Charles},
  title   = {The Proof and Measurement of Association between Two Things},
  journal = {The American Journal of Psychology},
  year    = {1904},
  volume  = {15},
  number  = {1},
  pages   = {72--101}
}

@book{Tango2007,
  author    = {Tango, Tom M. and Lichtman, Mitchel G. and Dolphin, Andrew E.},
  title     = {The Book: Playing the Percentages in Baseball},
  publisher = {Potomac Books},
  address   = {Washington, DC},
  year      = {2007}
}

@article{Albert2005,
  author  = {Albert, Jim},
  title   = {Does a Baseball Hitter's Batting Average Measure Ability or Luck?},
  journal = {STATS: The Magazine for Students of Statistics},
  year    = {2005},
  number  = {44},
  pages   = {3--6},
  note    = {Fall 2005/Winter 2006 issue}
}

@article{Bickel2004,
  author  = {Bickel, J. Eric},
  title   = {Why It's So Hard to Hit .400: New Insights into an Old Statistic},
  journal = {The Baseball Research Journal},
  year    = {2004},
  volume  = {32},
  pages   = {15--21}
}

\appendix
\section{Accuracy of the linear approximations}\label{app}

Approximation \eqref{eq:delta} replaces $\Var(\log f)$ by a first-order expansion of $\log\{h + (1-h)b\}$ about $(\bar h, \bar b)$. In 2025 the approximation is within \aGap{} of the exact value. Across the twelve decades from the 1900s to the 2010s the relative error has median \gapMed{} and maximum \gapMax{}. Approximation \eqref{eq:covsplit}, divided by $\bar c \bar f$, reproduces the exact 2010s value of $\Cov(\log c, \log f)$ to four decimal places (\covLogFull{} versus \covLogExact{}).

Noise corrections subtract the average binomial sampling variance from each observed variance: $\overline{p(1-p)/n}$ on the rate scale, with $n = AB$ for $c$, $AB - SO$ for $h$ and $AB - SO - HR$ for $b$, and $\overline{(1-p)/(np)}$ on the log scale. Covariances are not corrected, because the three rates are computed from disjoint sets of at-bats and their sampling errors are independent. ``Talent'' correlations in Tables~\ref{tab:logdecomp} and~\ref{tab:decades} divide the observed covariance by the noise-corrected standard deviations.

\end{document}